\documentstyle[prb,aps,twocolumn,epsf]{revtex}
\def\geqap{\,\raise 2pt \hbox{$>\kern-11pt \lower 5pt \hbox{$\sim$}$}\,}
\def\leqap{\,\raise 2pt \hbox{$<\kern-10pt \lower 5pt \hbox{$\sim$}$}\,}
\begin{document}
\draft
\title{Orbital Polarization and Fluctuation in Manganese Oxides}
\author{Ryo Maezono and  Naoto Nagaosa}
\address{Department of Applied Physics, University of Tokyo,
Bunkyo-ku, Tokyo 113-8656, Japan}
\date{\today}
\maketitle
%
\narrowtext
\section{Introduction}
Doped manganites ($R_{1-x}, A_x$)$_{n+1}$Mn$_n$O$_{3n+1}$ 
(R=La, Pr, Nd, Sm ; A= Ca, Sr, Ba ; $n=1, 2, \infty$)
have recently attracted considerable interests 
due to the colossal magnetoresistance (CMR) observed near the
ferromagnetic (spin-$F$-type) transition temperature $T_c$ 
\cite{chaha,helmolt,tokura95,jin94,ram,moritomo95-96}
It is now recognized that the most fundamental interaction in 
these materials is the double exchange interaction,
which connects the transport and magnetism. 
\cite{jonker50,zener51,anderson55,degenne60}
Since the discovery of the CMR\cite{tokura95}, however, it has been
pointed out that the double-exchange mechanism alone cannot explain 
not only the CMR \cite{millis95} but also the several observed 
properties in this system.
As the mechanism which plays an essential role on CMR in cooperate with 
double-exchange interaction,
several candidates has been suggested, for example, the Jahn-Teller (JT)
polaron \cite{millis95,roder96,millis96,alex99}, the charge inhomogeneity
\cite{yunoki98-1,yunoki98-2,yunoki98-3,moreo99},
the percolative processes \cite{jung99,cheong99}, 
the phase segregation with respect to
the orbital \cite{endoh99}, and the orbital polarization and 
fluctuation \cite{ishihara97,maezono98-1,maezono98-2}, 
the last of which we describe in this manuscript.
\par
The well-known issue showing the importance of the orbital polarization
is the layered ($A$-type) antiferromagnetism (AF) observed in
the mother compound of this system.\cite{wollan55,kugel72,ishihara96}
Kugel and Khomskii \cite{kugel72} treated this in a framework of the 
superexchange interaction and showed that the full consideration of the 
orbital degeneracy is indispensable to explain the spin $A$-type structure.
There the {\it orbital polarization} is essential:
Under the doubly degenerate orbitals, the on-site Coulombic repulsion 
differs depending on the configuration of the occupation,
$U$ for the two electrons occupying
the same orbital, $U'-J$ for occupying the different orbitals with the 
parallel spin, and $U'+J$ for occupying the different orbitals with the 
anti-parallel spins, where $U$ and $U'$ are the intra- and inter-orbital
Coulombic interactions, respectively, and $J$ is the interorbital exchange
interaction. \cite{kugel72}
In order to maximize the energy gain via the second-order 
perturbative processes, electrons form the staggered orbital occupation
(AF orbital ordering) with the energy gain by $t^2/(U'-J)$.\cite{kugel72}
In such an orbital ordering, there is a definite distinction between the 
occupied and unoccupied orbitals, which is the {\it orbital polarization}.
The orbital polarization (or orbital ordering) is the important origin of the
$A$-type spin structure in the mother compound.
In the viewpoint that the CMR with $x\sim 0.175$ \cite{tokura95}
occurs in the lightly doped Mott insulator, the orbital polarization
is likely to survive and to play an important role on CMR.
\par
Another point is that the origin of Hund's coupling $J_H$ is 
nothing but the on-site Coulomb interactions.
It seems therefore rather artificial to take $J_H \rightarrow \infty$ while
the on-site repulsion is neglected, as in the framework of the 
double-exchange mechanism. \cite{anderson55}.
\par
In a form which include the mother compound, 
we studied the extended Hubbard-type model with the orbital degeneracy
for any doping concentration.\cite{maezono98-1,maezono98-2}
Calculated meanfield phase diagram well reproduced the global topology 
of the magnetic structure depending on the doping concentration $x$.
\cite{maezono98-1,maezono98-2}
With a large orbital polarization we could predict the emergence of
the $A$-type and the rod-type ($C$-type) AF in the moderately doped
region, independently from the experiments discovering these phases
in the CMR compound with finite $x$ 
(which is larger than that for the CMR region)
\cite{kawano97,kuwahara98,moritomo98,kuwahara99,kajimoto99}
It turned out that these phases
couldn't be reproduced without a large orbital
polarization, which is therefore essential in the doped region where
these $AF$ phases are observed.
Because the orbital polarization increases as $x$ decreases, this 
concludes that the large orbital polarization survives even in the 
CMR region with smaller $x$.
Base on this result, we discuss the spin canting, the spin wave dispersion,
and spin wave stiffness from the standing point of the large orbital 
polarization.
We also discuss the orbital fluctuation which turned out to be 
important in the ferromagnetic metallic (FM) region where the CMR 
is observed.
Spin wave softening near the zone boundary is also discussed in this 
context.
\par
%
\section{Model and Formalism}
 We start with the Hamiltonian
\begin{equation}
H=H_K+H_{\rm Hund}+H_{\rm on\ site}+H_S
+H_{\rm el-ph}
 \ ,
\label{eqn:eq1}
\end{equation}
where $H_K$ is the kinetic energy of $e_g$ electrons,
$H_{\rm Hund}$ is the Hund's coupling between $e_g$ and 
$t_{2g}$ spins, and $H_{\rm on\ site}$ represents the on-site
Coulomb interactions between $e_g$ electrons.
$t_{2g}$ spins are treated as the localized spins with $S=3/2$.
The AF coupling between nearest neighboring $t_{2g}$ spins 
is introduced in $H_S$ to reproduce the NaCl-type ($G$-type) AF spin ordering 
observed at $x=1.0$. \cite{wollan55}
Using an operator $d_{i\sigma \gamma }^{\dagger}$ which creates
an $e_g$ electron with spin $\sigma$ ($={\uparrow}, {\downarrow}$)
in the  orbital $\gamma$ [$=a(d_{x^2-y^2}), b(d_{3z^2-r^2})$] at site $i$,
each term of Eq. (\ref{eqn:eq1}) is given as 
\begin{equation}
H_K=
\sum\limits_{\sigma \gamma \gamma' \langle ij \rangle} 
{t_{ij}^{\gamma \gamma '}d_{i\sigma \gamma }^{\dagger}d_{j\sigma \gamma '}}\ ,
\label{eqn : eq2}
\end{equation}
\begin{eqnarray}
H_{\rm Hund}=-J_H\sum\limits_i {\vec S_{t_{2g} i}\!\cdot\! \vec 
S_{e_g i}}\ ,
\label{eqn : eq4}
\end{eqnarray}
and 
\begin{equation}
H_S=J_S\sum\limits_{\left\langle {ij} \right\rangle } {\vec S_{t_{2g} i}
\!\cdot\! \vec S_{t_{2g} j}}\ .
\label{eqn : eq3}
\end{equation}
$t_{ij}^{\gamma \gamma '}$ in $H_K$ is the 
electron transfer integral between nearest neighboring sites 
and it depends on the pair of orbitals and the direction of 
the bond as follows: \cite{ishihara96} 
\begin{equation}
  t_{i \ i+x}^{\gamma \gamma '}= t_0\left( {\matrix{{-{3 \over 4}}
     &{{{\sqrt 3} \over 4}}\cr
     {{{\sqrt 3} \over 4}}&{-{1 \over 4}}\cr
     }} \right) \ ,
     \\ 
\end{equation}
\begin{equation}
  t_{i \ i+y}^{\gamma \gamma '} = t_0\left( {\matrix{{-{3 \over 4}}
     &{-{{\sqrt 3} \over 4}}\cr
     {-{{\sqrt 3} \over 4}}&{-{1 \over 4}}\cr
     }} \right) \ ,
     \label{eqn : eq51}  
\end{equation}
and 
\begin{equation}
      t_{i \ i+z}^{\gamma \gamma '} = t_0\left( {\matrix{0&0\cr0&{-1}\cr
     }} \right) \ . 
     \end{equation}
$t_0$ is the electron transfer integral between $d_{3z^2-r^2}$ orbitals 
along the $z$ direction. 
The spin operator for the $e_g$ electron is defined as 
$\vec S_{e_g i}={1 \over 2}\sum\limits_{\gamma \alpha \beta} 
 {d_{i\gamma \alpha }^{\dagger}\vec \sigma _{\alpha \beta }
d_{i\gamma \beta }}$ with the Pauli matrices $\vec \sigma$. 
$\vec S_{t_{2g}i}$ denotes the localized $t_{2g}$ spin on the 
$i$ site with $S=3/2$.
$H_{\rm on\ site}$ consists of 
the intra- and the inter-orbital Coulomb interactions and 
the inter-orbital exchange interaction;
\cite{maezono98-2,ishihara96}
\begin{eqnarray}
H_{\rm on\ site}
&=& U\sum\limits_{j\gamma } 
{n_{j\gamma \uparrow }}n_{j\gamma \downarrow }  \nonumber \\
&\ \ \ \ &+ U'\sum\limits_{j\sigma \sigma '} {n_{ja\sigma }}n_{jb\sigma '} 
 \nonumber \\
&\ \ \ \ &+ J\sum\limits_{j\sigma \sigma '} {d_{
ja\sigma }^{\dagger}d_{jb\sigma '}^{\dagger}d_{ja\sigma '}
d_{jb\sigma }}  \nonumber \\
&=& -\sum\limits_i {\left( {\tilde \beta \vec T_i^2+\tilde
 \alpha \vec S_{e_{g} i}^2} \right)} \ , 
\label{eqn : eq5}
\end{eqnarray}
where $n_{j\gamma \sigma }=d_{j\gamma\sigma }^{\dagger}
d_{j\gamma\sigma}$ and $n_{j\gamma }=\sum\limits_\sigma
  {n_{j\gamma \sigma }}$, and the isospin operator describing 
the orbital degrees of freedom is defined as 
\begin{equation}
\vec T_i={1 \over 2}\sum\limits_{\gamma \gamma' \sigma}  {d_{i\gamma \sigma }
^\dagger\vec \sigma _{\gamma \gamma '}d_{i\gamma '\sigma }} \ .
\label{eqn : eq9}
\end{equation}
Coefficients of the spin and isospin operators, i.e., 
$\tilde \alpha $ and $\tilde \beta $, are given as
\cite{maezono98-2,ishihara96}
\begin{equation}
\tilde \alpha = U-{J \over 2}>0\ , 
\end{equation}
and 
\begin{equation}
\tilde \beta = U-{3J \over 2}>0 \ .
\label{eqn : eq11}
\end{equation}
The magnitude of the meanfield solution of the isospin operator 
$\langle\vec T\rangle$ gives the energy splitting between
the occupied and unoccupied orbitals, namely the {\it orbital polarization}.
Therefore, the minus sign in Eq. (\ref{eqn : eq2}) means that
the on-site repulsion in this system induces not only the spin polarization 
but also the orbital polarization as the interplay with the 
orbital degree of freedom.
By this orbital polarization, the anisotropy of the $e_g$ orbitals is fully
reflected to the transport and hence introduces the lower dimensionality
even in the system with the isotropic crystal structure 
($n=\infty$ ; 113-system).
The parameters $\tilde \alpha ,\tilde \beta ,
t_0$, used in the numerical calculation are chosen as
$t_0 = 0.72$ eV, $U=6.3$ eV, and $J=1.0$ eV, being 
relevant to the actual manganese oxides. \cite{maezono98-1,maezono98-2} 
The electron-phonon interaction is given as,\cite{maezono98-2}
\begin{eqnarray}
    H_{\rm el-ph}
    = +\left|g\right|r\sum_{i}{\vec v_{i}\cdot\vec T_{i}}\ ,
\label{eqn : eq3.2.16}
\end{eqnarray}
where $g$ is the coupling constant and $r$ ($\vec v_{i}$) is
the magnitude (direction) of the lattice distortion of the MnO$_6$-octahedra.
Values of $r$ and $\vec v$ are taken from the observed elongation as,
$r\sim0.028$ and $\vec v=\pm(\sqrt{3}/2)\hat x-(1/2)\hat z$ 
(staggered as $d_{3x^2-r^2}/d_{3y^2-r^2}$) for
LaMnO$_3$ ($n=\infty$),\cite{matsumoto70} and $r\sim0.01$, $\vec v /\!/ \hat z$
(elongation along $c$-axis) in (La$_{1-x}$, Sr$_x$)$_3$Mn$_2$O$_7$ 
($0.3<x<0.4$). \cite{moritomo98}
\par
In the path-integral representation,
the grand partition function is represented as 
\begin{equation}
\Xi =\int { \prod\limits_{i}
D\vec S_{t_{2g} i}}D\bar d_{i\gamma \sigma }
  Dd_{i\gamma \sigma }\exp \left\{ {-\int {d\tau \kern 
  1pt L\left( \tau  \right)}} \right\} \ ,
  \label{eqn : eq12} 
\end{equation}
with
\begin{equation}
  L\left( \tau  \right)=H\left( \tau  \right)+\sum\limits_
  {\sigma \gamma i} {\bar d_{i\gamma\sigma}\left( \tau  
  \right)\left( {\partial _\tau -\mu } \right)d_{i\gamma\sigma}}
\left( \tau  \right)$$ \ ,
\label{eqn : eq13}
 \end{equation}
where $\tau$ is the imaginary time
introduced in the path-integral formalism, and
$ {\bar d_{i\gamma \sigma }} $, $d_{i\gamma \sigma }$ 
are the Grassmann variables corresponding to the
operators $ d_{i \gamma \sigma }^{\dagger}$ and $d_{i\gamma 
\sigma } $, respectively. 
By introducing two kinds of auxiliary fields corresponding 
to the following mean-field solutions,\cite{maezono98-2}
\begin{eqnarray}
    \vec \varphi_S
   & = & \left\langle \vec S_{e_g} \right\rangle + {{J_H } \over 
    {2 \tilde \alpha }} \left\langle \vec S_{t_{2g}} \right\rangle \ ,\\
    \vec \varphi_T
   & = & \left\langle \vec T \right\rangle \ ,
\label{eqn : eq3.2.16}
\end{eqnarray}
we obtain the effective action with respect to these auxiliary fields
and $\vec S_{t_{2g}}$, after integrating over the fermion variables 
as,\cite{maezono98-2}
\begin{eqnarray}
 \Xi & \!=\! & \int {D\left\{ \varphi  \right\}} e^{S_{e\!f\!f}[\vec\varphi]}
 \ ,
 \label{eqn : eq1.2.3a}  
\end{eqnarray}
\begin{eqnarray}
 S_{\rm eff}[\vec\varphi]
   & = & {Tr\ln G_{kk';nn';\gamma \gamma ';\alpha 
    \beta }^{-1}-\int {d\tau \kern 1pt L_{\vec \varphi}}} \ ,
\label{eqn : eq1.2.3}  
\end{eqnarray}
\begin{eqnarray}
   L_{\vec \varphi} & =  &  J_S \sum\limits_{\left\langle {ij} \right\rangle } 
          {\vec S_{t_{2g} i}(\tau)\cdot \vec S_{t_{2g} j}(\tau)}
          - J_H\sum\limits_i {\vec S_{t_{2g} i}(\tau)} 
            \cdot \vec\varphi_{S i}(\tau) \nonumber \\
    &   & + \tilde \alpha \sum\limits_i{\vec \varphi_{S i}^2(\tau)}
          + \tilde \beta \sum\limits_i {\vec \varphi_{T i}^2(\tau)}  \ ,
\label{eqn : eq1.2.4}    
\end{eqnarray}
\begin{eqnarray}
 \lefteqn{ G_{kk';nn';\gamma \gamma ';\alpha 
    \beta }^{-1}  } \cr
   &\ \ \ \  &=  { \left( {-i\omega _n-\mu } \right) 
   \delta_{kk';nn';\gamma \gamma ';\alpha \beta }
       \!+\! M_{kk';nn';\gamma \gamma ';\alpha \beta } }\ ,
  \label{eqn : eq1.2.6}    
\end{eqnarray}
\begin{eqnarray}
  \lefteqn{M_{kk';nn';\gamma \gamma ';\alpha \beta }}\cr
& \ \ \ \  & =
   \varepsilon _k^{\gamma \gamma '}\delta _{kk'}\delta 
    _{nn'}\delta _{\alpha \beta } \cr
& \ \ \ \  &  \ \ \ \ 
 - {{\tilde \alpha } \over 
    {\sqrt {\beta N}}} \vec \sigma _{\alpha \beta } \cdot \vec\varphi 
    _S(k-k',\omega_n-\omega_{n'})\delta _{\gamma \gamma '} \cr
& \ \ \ \  & \ \ \ \ 
 - {{\tilde \beta } \over {\sqrt {\beta N}}} \vec \sigma
    _{\gamma \gamma '} \cdot \vec \varphi _T(k-k',\omega_n-\omega_{n'})
    \delta _{\alpha \beta }  \ ,
\label{eqn : eq1.2.7}
\end{eqnarray}
where we have introduced the momentum representation,
\begin{equation}
  \varphi _{x j}(\tau)  = {1 \over {\sqrt {\beta N}}}\sum\limits_k 
    {\sum\limits_{n} {\varphi _x(k,\omega_n)}}e^{i\vec k
    \cdot \vec R_j-i\omega _n\tau } \ ,
\label{eqn : eqN.5}
\end{equation}
for $x=S,T$.
\par
In the meanfield approximation, the free energy is given as
\begin{eqnarray}
 F_{MF} & \!=\! & -k_B T \cdot S_{e\!f\!f}[\vec\varphi^c]+\mu N
 \ ,
 \label{eqn : eq1.2.3a}  
\end{eqnarray}
where $\vec\varphi^c$ denotes the saddle point of $\vec\varphi_{S,T}$.
We seek the saddle point within the several assumed 
ordering configurations, as following:
We consider four kinds of the spin alignment in the cubic cell: spin $F$, 
$A$, $C$ and $G$ (NaCl-type).
For spin $A$, we also consider the possibility of the canting characterized
by an angle $\eta$ which is $0\ (\pi)$ for spin $F$ ($A$).
As for the double-layered compounds ($n=2$), we consider an isolated 
double-layer, for which the Brillouin zone contains only two 
$\vec k$-points along $c$-axis, because the exchange interaction 
between two double-layers
is reported to be less than 1/100 compared with the intra double-layer one. 
\cite{fujioka99}
As for the orbital degrees of freedom, we consider two sublattices $I$,
and $I\!I$, on each of which the orbital is specified by the
angle $\theta_{I,I\!I}$ as \cite{maezono98-2}
\begin{eqnarray}
\left| {\theta _{I,I\!I}} \right\rangle =\cos {{\theta _{I,I\!I}} 
\over 2}\left| {d_{x^2-y^2}} \right\rangle +\sin {{\theta _{I,I\!I}} 
\over 2}\left| {d_{3z^2-r^2}} \right\rangle.
\label{eqn : eqN.3}    
\end{eqnarray}
We also consider four types of orbital-sublattice ordering, i.e., 
$F$-, $A$-, $C$-, $G$-type in the cubic cell.
Henceforth, we often use a notation such as spin A, orbital $G$
($\theta_I,\theta_{I\!I}$) etc..
Denoting the wave vector of the spin (orbital) ordering as
$\vec q_{S}$ ($\vec q_{T}$), the ground state energy is
given as a function of the spin ordering ($\eta$, $\vec q_{S}$),
the orbital ordering ($\theta_{I,I\!I}$, $\vec q_{T}$), and
the lattice distortion ($g$, $r$, $\vec v$).
\par
In the random-phase-approximation (RPA), we expand $S_{e\!f\!f}[\vec\varphi]$
with respect to the small fluctuation $\delta\vec\varphi_S$
from its mean-field solution $\vec \varphi_S^c$ for the spin
degrees of freedom,
\begin{eqnarray}
    \vec \varphi_S
   & = & \vec \varphi_S^c + \delta\vec\varphi_S \ .
\label{eqn : eq3.2.16c}
\end{eqnarray}
Denoting the perpendicular (parallel) component to the
mean-field as $\vec\pi$ ($\vec \sigma$),
\begin{eqnarray}
 \delta \vec \varphi _S(k,\omega_n)
=\vec \sigma(k,\omega_n)+\vec \pi(k,\omega_n) \ ,
\label{eqn : eq1.3.16b}
\end{eqnarray}
the deviation of the action can be written as \cite{maezono99-2}
\begin{eqnarray}
\lefteqn{
\delta S_{\rm eff}=
\sum\limits_{q,\Omega } {K_\pi \left( {\vec q,\Omega } \right)
 \pi \left( {\vec q_S\!+\! \vec q,\Omega } \right)
\!\cdot\!
\pi }
\left( {-\vec q_S\!-\!\vec q,-\Omega } \right)
} \cr
&\ \ \ \ & \!+\!\sum\limits_{q,\Omega }\! {K_{\!\times\!}\! 
\left( {\vec q,\Omega } \right) \vec \pi \!
\left( {\vec q_S\!+\!\vec q,\Omega } \right)
\!\cdot\! \left\{ {\vec n\!\times\! \vec \pi 
\left( {-\vec q_S\!-\!\vec q,-\Omega } \right)} \right\}}.
\label{eqn : eq3}    
\end{eqnarray}
Because the spin wave is the Goldstone boson, the condition
$
K_\pi \left( {0,0} \right)=0 \ , 
K_{\!\times\!} \left( {0,0} \right)= 0 ,
$
can be derived.
Coefficient of the diagonalized quadratic form is obtained as
$
K_{\uparrow \left( \downarrow  \right)}
=K_\pi \pm iK_\times \ ,
$
zero-point of which $\left(K_{\uparrow \left( \downarrow  \right)}
(\vec q,\Omega=-i\omega)=0\right)$ gives the dispersion relation of 
the excitation $\omega=\omega(\vec q)$. 
$K_{\sigma}(\vec q,\Omega)$ can be expanded as, 
\begin{eqnarray}
\lefteqn{{{K_{\sigma}\left( {\vec q,\Omega } \right)} \over {\tilde \alpha }}
}\cr &\ \ \ \ &
=\left\{ {\matrix{{\sigma A\cdot i\Omega +\sum\limits_{\alpha =x,y,z} 
{C_\alpha q_\alpha ^2}\ \ \ \ \cdots Spin\ F\ \ \ \ }\cr
{B\Omega ^2+\sum\limits_{\alpha =x,y,z} {C_\alpha q_\alpha ^2}
\ \ \ \cdots Spin\ AF}\cr
}} \right. \ ,
\label{eqn : eqN.1}
\end{eqnarray}
where $\sigma=1\ (-1)$ corresponds to spin up (down), respectively. 
We evaluate only the {\it static} spin-wave stiffness 
$C_\alpha = C_\alpha(x)$ because the {\it dynamic} spin wave velocity
evaluated by using the above expression inherently gives a 
misleading estimation;
For the half-filled insulator, $x=0$, it does not reduce to the energy 
order as the superexchange interaction $\sim t^2/U$, giving rather
the order of $t$,\cite{fradkin91} perhaps due to the inherent fault of the RPA.
For the metallic region, $x\ne 0$, 
we cannot reproduce the correct dispersion-relation, because 
the Landau-damping is not properly treated in our calculation where
the Brillouin zone is discretized and thus the gapless 
individual-excitation is not correctly evaluated.
$C_\alpha = C_\alpha(x)$ roughly reflects the exchange-interaction 
depending on $x$.
\par
%
%
\section{Orbital Polarization and Fluctuation}
\begin{figure*}
\begin{center}
\vspace{0mm}
\hspace{0mm}
\epsfxsize=8cm
\epsfbox{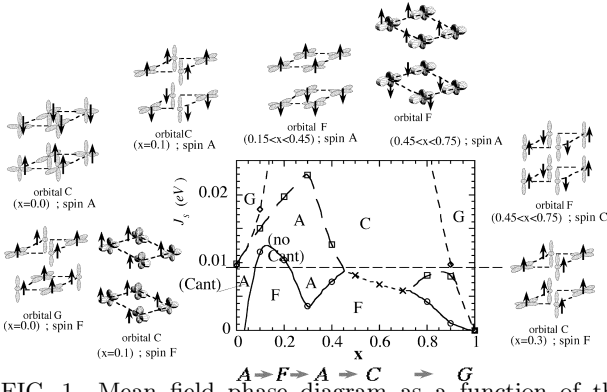}
\vspace{0mm}
\caption[aaa]{Mean field phase diagram as a function of the carrier 
concentration ($x$) and the antiferromagnetic interaction 
between $t_{2g}$ spins ($J_S$) for the cubic system 
($n=\infty$ ; 113-system).\cite{maezono98-1,maezono98-2}
Dotted line ($J_S$ =0.009) well reproduces the change of the spin 
structure experimentally observed.}
\label{fig : F1}
\end{center}
\end{figure*}
Fig. \ref{fig : F1} shows the zero-temperature meanfield phase diagram 
of the cubic system ($n=\infty$ ; 113-system) in a plane of $x$ and 
$J_S$ (AF superexchange interaction between $t_{2g}$ spins), 
with the optimization of the orbital at each point on the plane. 
\cite{maezono98-1,maezono98-2}
With $J_S$ being fixed to be a relevant value to the actual compounds,
$J_S$ =0.009, \cite{ishihara96,goodenough55} we obtain
the spin transition as $A\!\rightarrow\!F\!\rightarrow\!A\!\rightarrow\!C\!
\rightarrow\!G$ with increasing $x$, being consistent with experiments
\cite{kuwahara98}.
Non-monotonic phase boundaries are essential for these variety of the 
spin structures.
\begin{figure}
\begin{center}
\vspace{0mm}
\hspace{0mm}
\epsfxsize=8cm
\epsfbox{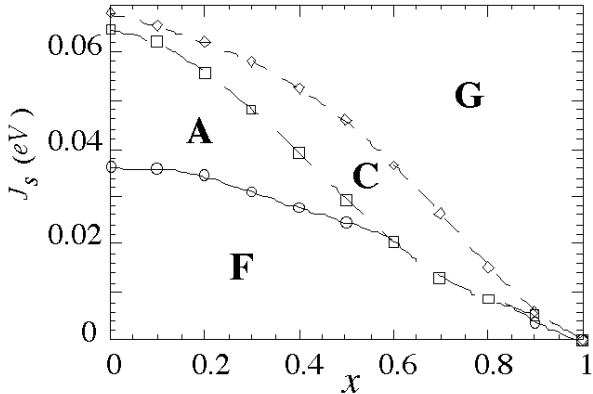}
\vspace{0mm}
\caption[aaa]{Mean field phase diagram with no orbital polarization.
In this case the nonmonotonic behavior of the phase boundaries 
disappears.}
\label{fig : F1-2}
\end{center}
\end{figure}
\noindent
Dimensionality control by the orbital polarization is the origin
of such a behavior:
Orbital ordering changes from that maximizing the superexchange 
energy gain for smaller $x$ ($\le 0.3$) to that maximizing the
double-exchange energy gain for larger $x$, with the change
in the dimensionality. \cite{maezono98-2}
This orbital transition varies the kinetic energy gain non-monotonically
via the change in the density of states with the van-Hove singularity.
\cite{maezono98-2}
This can also be an origin of the instability toward the phase segregation, 
\cite{yunoki98-1,yunoki98-2,yunoki98-3,moreo99,endoh99}
though it does not occur in our calculation.
\par
With no orbital polarization, such a non-monotinic behavior cannot
be reproduced, \cite{maezono98-2}
as shown in Fig. \ref{fig : F1-2}.
This is because the anisotropy of the degenerate orbitals are mixed
to disappear with no polarization (in this case we cannot say which
orbital is occupied because of the hybridization).
\begin{figure}[p]
\begin{center}
\vspace{0mm}
\hspace{0mm}
\epsfxsize=8cm
\epsfbox{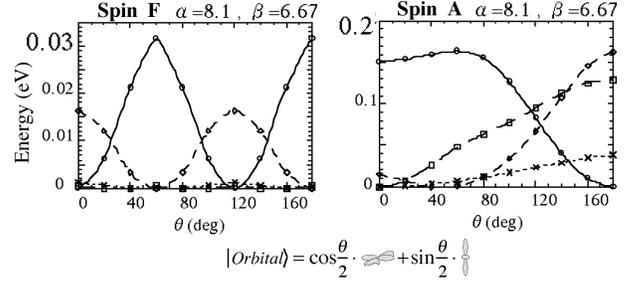}
\vspace{0mm}
\caption[aaa]{The energy as a function of the orbital state characterized 
by $\theta$ in the several value of $x$. (a) Spin $F$ is assumed. 
(b) Spin $A$ is assumed. 
In both cases, the orbital $F$-type structure is assumed. \cite{maezono98-2}}
\label{fig : F2}
\end{center}
\end{figure}
\noindent
For the global topology $A\!\rightarrow\!F\!\rightarrow\!A\!\rightarrow\!C\!
\rightarrow\!G$ to be reproduced, it is therefore necessary that
a large orbital polarization occurs even in the spin $F$ phase where 
CMR is observed.
As for the spin $F$ (CMR) phase, however due to its isotropy,
the question is how does it coexist with the observed isotropic
properties in CMR compounds, \cite{tokura95,martin96}
because such a polarization leads to the anisotropic carrier hopping.
The key for this question is the orbital fluctuation. 
\par
Fig. \ref{fig : F2} shows the energy dependence on the orbital 
configuration for spin $F$ and $A$ phases. \cite{maezono98-2}
In spin $F$ phase, there are many degenerate saddle points due to the 
isotropy, and the height of the barrier is an order smaller than that 
for the $AF$ phase. 
\begin{figure}[p]
\begin{center}
\vspace{0mm}
\hspace{0mm}
\epsfxsize=8cm
\epsfbox{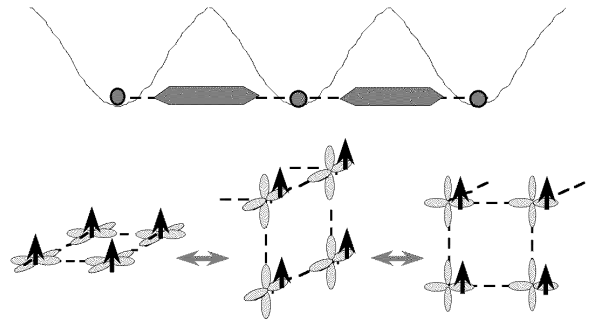}
\vspace{0mm}
\caption[aaa]{Schematic picture of the orbital liquid state.}
\label{fig : F2-2}
\end{center}
\end{figure}
\noindent
This result implies that the orbital fluctuation becomes important in the 
spin $F$ phase comparing with the $AF$ phases. 
Reentrant of the spin $A$ with increasing $x$, seen in Fig. \ref{fig : F1}
implies that the $d_{x^2-y^2}$ orbital ordering is inherent property of the
double-exchange interaction.
Therefore, in the extent beyond the mean field theory, it is likely that 
the degenerate saddle points, $d_{x^2-y^2}$, $d_{y^2-z^2}$, and $d_{z^2-x^2}$
resonate to recover the isotropy of the spin $F$ metallic phase
though the large orbital polarization still survives, forming
the \it{orbital liquid}\rm\ state, as shown in Fig. \ref{fig : F2-2}.
\cite{ishihara97}
%
\section{Spin canting and Orbital ordering}
\begin{figure}[p]
\begin{center}
\vspace{0mm}
\hspace{0mm}
\epsfxsize=8cm
\epsfbox{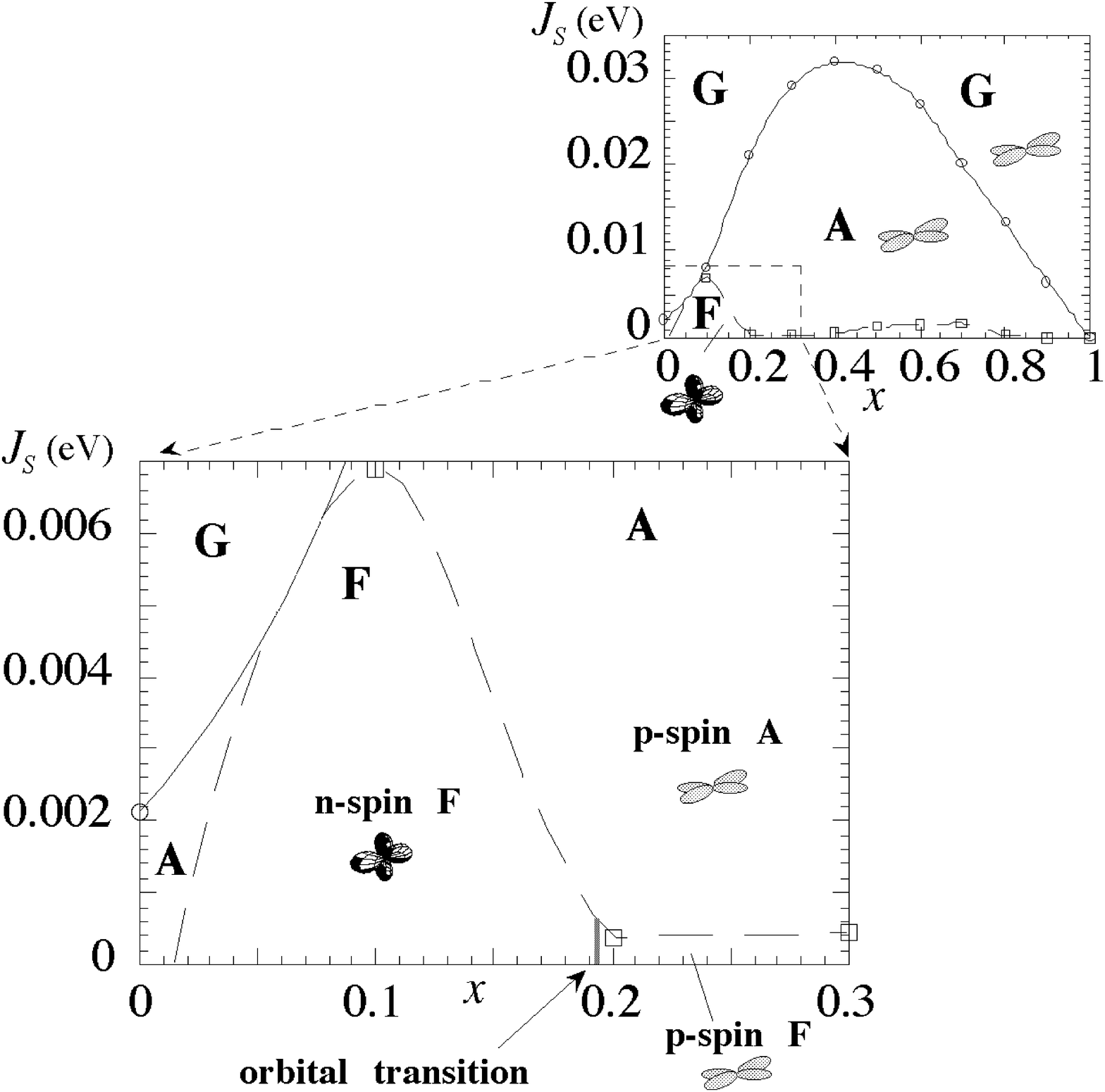}
\vspace{0mm}
\caption[aaa]{Phase diagram for the layered 
compounds ($n=2$ ; 327-system). \cite{maezono99-2}}
\label{fig : F3}
\end{center}
\end{figure}

Fig. \ref{fig : F3} shows the phase diagram of the layered 
compounds ($n=2$ ; 327-system). \cite{maezono99-2}
In this system, the anisotropic crystal structure also controls the
dimensionality:
restricted hopping along $c$-axis brings about the 
$d_{x^2-y^2}$-orbital ordering in the metallic region even for the isotropic
spin $G$- and $F$ ($x>0.2$) alignment.
Especially the planer spin $F$ phase seen for $x>0.2$ is essential for 
the spin canting observed in this system 
\cite{argyriou971,argyriou972,perring97,hirota98,kimura98,kubota98}
with $0.4 \leq x_{exp.} \leq 0.48$, as below.
The global topology of the phase diagram in this system is reproduced
as $A\!\rightarrow\!F\!\rightarrow\!A\!\rightarrow\!G$ with increasing $x$,
\cite{mitchell99} 
as shown in Fig. \ref{fig : F3}, where there are two phase boundaries 
between the spin $A$ and $F$ phases; One is with small $x$ ($x<0.1$,
{\it left} boundary), and the other is with finite $x$ ($x>0.1$,
{\it right} boundary). 
Under the competition between the super- and double-exchange interaction, 
the saddle point of the canting angle $\eta$ is given as,\cite{maezono99-2}
\begin{equation}
\cos{\frac{\bar\eta}{2}} = \frac{t_z x}{4J_S} \ ,
\label{eqn : cond}
\end{equation}
where the $t_z$ denotes the inter-layer 
hopping integral.
The right-hand-side of the above equation should be
smaller than the unity for the occurrence of the spin canting.
For the {\it left} boundary, this condition can be satisfied 
for any magnitude of $t_z$ because the small $x$ always makes 
the right-hand-side of Eq. (\ref{eqn : cond}) to be small. 
For the {\it right} boundary, however, this can be satisfied 
only when the orbital is planer, i.e., nearly $d_{x^2-y^2}$ 
with small $t_z$;
the characteristic energy scale of the hopping integral is an order 
greater than 
that for $J_S$, therefore if the orbital is spherical,
$t_z/J_S$ becomes much more than the unity and hence the 
right-hand-side of Eq. (\ref{eqn : cond}) because in this case
the finite $x$ does not make it small any more.
Therefore it is concluded that the planer orbital is indispensable 
for the canting observed on the {\it right} phase boundary with finite $x$
(metallic canting).
\par
Experimentally, this metallic canting is commonly found in the double-layered
compound ($n=2$).
\cite{argyriou971,argyriou972,perring97,hirota98,kimura98,kubota98}
This is because the layered structure stabilizes the planer 
($d_{x^2-y^2}$) orbital in the metallic region.
In the 113-system, on the other hands, no spin canting on the 
{\it right} boundary is reported, \cite{kawano97} 
which may be accounted by its isotropy
leading to no such stabilization.
This isotropy of the 113-compounds may allow only two possibilities for 
its orbital state; one is the orbital liquid state resonating among
the planer orbitals, $d_{x^2-y^2}$, $d_{y^2-z^2}$, and $d_{z^2-x^2}$, 
\cite{ishihara97} and the other is the quasi-spherical orbital, 
which is obtained as the saddle point within the extent of the 
meanfield theory, as shown in Fig. \ref{fig : F1}. 
\cite{maezono98-1,maezono98-2}
Kawano \it{et al.} \rm observed the metallic canting in 113-system,
Nd$_{1/2}$Sr$_{1/2}$MnO$_3$, with slight anisotropy of the lattice 
structure \cite{kawano98} in the temperature-driven transition
between the spin $F$ (high-temperature phase) and the spin $A$
(low-temperature phase).
This supports the former possibility of the orbital, i.e., the
orbital liquid state;
If the orbital is quasi-spherical in perfectly cubic system, 
taking the latter possibility, such a slight lattice anisotropy 
leads to only a slight distortion of the spherical orbital which 
remains the right-hand-side of Eq. (\ref{eqn : cond}) still larger than the
unity and hence no canting is expected. 
On the other hand, taking the former possibility, such a slight anisotropy 
is enough to stabilize $d_{x^2-y^2}$ immediately and hence the metallic 
canting can be explained.
\par
Another important feature as for the metallic canting is the stability
of the spin $A$ phase against the canting.
When the $x$ holes are introduced, the kinetic energy gain 
$\Delta E_{kin}(\xi)$ via the bonding and anti-bonding splitting 
$\Delta=t_z\cos{\frac{\eta}{2}=t_z \xi}$.\cite{anderson55} 
is given as, \cite{maezono99-2}
\begin{eqnarray}
\Delta E_{kin}\left(\xi\right) 
\!\sim\!
\left\{
\begin{array}{ll}
\!\!-t_z^2\!\cdot\!N_F\cdot \xi^2 
\ \ (\rm for\ \it \xi < \xi_c \equiv \frac{x}{N_Ft_z}) 
\\
-t_z\cdot x\cdot \xi \ \ (\rm for\ \it \xi > \xi_c) 
\\
\end{array}
\right.
\ ,
\label{eqn : eq6.5.1}  
\end{eqnarray}
with simplifications of a perfect spin polarization and the
constant density of states.
The competition between this kinetic energy gain and the energy cost of the 
exchange interaction, $J_S\!\cos{\eta}=J_S\left(2\xi^2-1\right)$, is 
the origin of the spin canting. 
The lower line of Eq. (\ref{eqn : eq6.5.1}) is obtained by
de Gennes, and if this holds the canting always occurs.\cite{degenne60}
The new aspect here is that $\Delta E_{kin}(\xi) \propto \xi^2$
when the splitting $\Delta=t_z \xi$ is smaller than the Fermi energy
$\epsilon_F = x/N_F$ and both the bonding and anti-bonding bands are occupied.
Therefore the spin $A$ structure ($\xi=0,\ \eta=\pi$) is at least locally
stable when $2J_S > t_z^2 N_F$.
This condition can be satisfied when the orbital is almost $d_{x^2-y^2}$
and $t_z$ is much reduced from $t_0$.
By minimizing the total energy
$\Delta E (\xi) = \Delta E_{kin}(\xi)+\Delta E_{ex}(\xi)$,
it is found that the spin canting can occur only when $\xi_c < 1$;
When $\xi_c > 1$ ($x > t_z N_F$), only the upper line of 
Eq. (\ref{eqn : eq6.5.1}) is relevant and $\Delta E = \left(2J_S
-t_z^2 N_F\right)\cdot \xi^2$.
Therefore $\xi$ jumps from 1 (spin $F$) to 0 (spin $A$) as $J_S$
increases across $t_z^2 N_F/2$.
\begin{figure*}
\begin{center}
\vspace{0mm}
\hspace{0mm}
\epsfxsize=8cm
\epsfbox{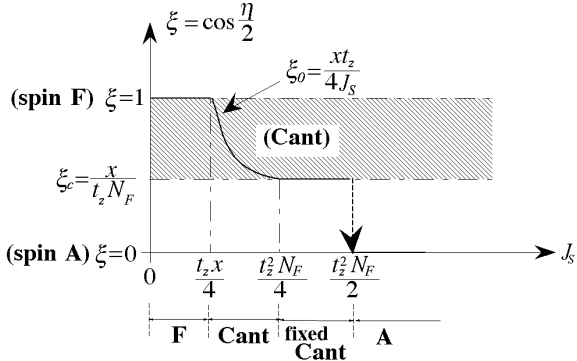}
\vspace{0mm}
\caption[aaa]{the optimized $\xi$ as a function of $J_S$ for
the case $\xi_c<1$. \cite{maezono99-2}}
\label{fig : F6}
\end{center}
\end{figure*}
When $\xi_c<1$ (the spin canting can occur), 
the optimized $\xi$ as a function of $J_S$ is given in 
Fig. \ref{fig : F6}.
As $J_S$ increases, the spin structure changes as spin $F$ 
($J_S < t_z x/4$) $\rightarrow$ spin canting ($t_z x/4 <J_S<t_z^2 N_F/4$)
$\rightarrow$ spin canting with fixed canting angle 
($t_z^2 N_F/4<J_S<t_z^2 N_F/2$) $\rightarrow$ spin $A$ ($t_z^2 N_F/2<J_S$).
Note that the canting angle continuously evolves from spin $F$, but jumps
at the transition to the spin $A$.
This seems to be consistent with experiments \cite{kubota98} where
the canting angle larger than 63 deg. is not observed.
\par
%
\section{Spin dynamics and Orbital}
\begin{figure}[htbp]
\begin{center}
\vspace{0mm}
\hspace{0mm}
\epsfxsize=8cm
\epsfbox{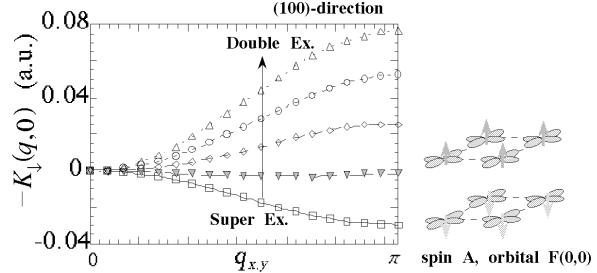}
\vspace{0mm}
\caption[aaa]{wave vector dependence of $-K_\downarrow\left(q_x,0\right)$
calculated for the spin $A$, $d_{x^2-y^2}$ orbital ordering, as an example.
Minus sign of $K_{\downarrow}$ comes from the negative $B$ in 
Eq. (\ref{eqn : eqN.1}) to correspond the positive sign of the plot 
with the stability of the saddle point.}
\label{fig : mF5}
\end{center}
\end{figure}

By fitting $K_\sigma\left(\vec q,0\right)$ as a function of $\vec q$, in
Eq. (\ref{eqn : eqN.1}), we can evaluate the {\it static} stiffness of the 
spin wave excitation $C_\alpha$ due to the $e_g$ orbital contribution.
As the orbital configurations to be assumed,
we take the saddle-point solution obtained in the meanfield theory 
as, \cite{maezono98-1,maezono98-2}
\begin{tabbing}
xxx \= $x=0.5-0.999$  \= Spin CCC \=  Orbital F:(180) \kill
\>$x=0.0$ \> Spin A \> Orbital C:($60,-60$) \\
\>$x=0.1$ \> Spin F \> Orbital C:($80,-80$) \\
\>$x=0.2-0.4$ \> Spin A \> Orbital F:($0$,$0$) \\
\>$x=0.5-0.9$ \> Spin C \> Orbital F:($180$,$180$). \\
\label{tbl : saddle point}
\end{tabbing}
\noindent
As for $x=0$, we further introduced the JT effect \cite{maezono98-2} by putting
the observed distortion of the MnO$_6$ octahedra.\cite{matsumoto70}
Fig. \ref{fig : mF5} shows the $q$-dependence of 
$-K_{\downarrow} \left(q_x,0\right)$ for the spin $A$ configuration
with $d_{x^2-y^2}$ orbital ordering (Minus sign of $K_{\downarrow}$ comes 
from the negative $B$ in Eq. (\ref{eqn : eqN.1}) to correspond the positive 
sign of the plot with the stability of the saddle point).
We have chosen this structure, as an example, because the double exchange 
interaction is most effective in this ordering as the meanfield theory shows, 
\cite{maezono98-2} and hence the crossover from super- to double-exchange
manifests itself most remarkably.
The enhancement of the stiffness with increasing $x$ can be reproduced.
This is due to the crossover from the super- to the double-exchange 
interaction as $x$ increases, which is well evaluated in our formalism 
in the unified way, as the inter- and intra-band transitions, respectively.
Plots are well fitted in the whole Brillouin zone for the 
ferromagnetic-bond direction by
\begin{eqnarray}
-K_{\downarrow}\left(\vec q, 0 \right)
\propto \left(1- \cos{q_\alpha}\right),
\label{eqn : eqN.3}    
\end{eqnarray}
not only in this case but also for all the other ordering
shown in the above table.
This implies that only the nearest-neighbor interactions
are important in the spin wave excitation.
This issue is important because there is no guarantee that the
exchange interaction can be represented by the nearest neighbor
Heisenberg model at finite doping, and because the softening
near the zone boundary has been observed in some materials
experimentally.\cite{hwang98,fernandez98,dai99,lynn99}
Our result here is in sharp contrast to the first principle study
\cite{solovyev96} which attributes the origin of such a softening to 
the longer-range interactions than the nearest-neighbor interactions.
Negative stiffness $C_\alpha<0$ seen for $x=0$, (100)-direction,
corresponds to the instability of 
the spin structure, which can be explained as follows.
Around $x\!=\!0$ the spin structure is dominated by the superexchange
interaction where the energy gain for spin-$F$ ($AF$) bond is
${{t_{o-u}^2} \mathord{\left/ {\vphantom {{t_{o-u}^2} 
\tilde\beta }} \right. \kern-\nulldelimiterspace} \tilde\beta }$
(${{t_{o-o}^2} \mathord{\left/ {\vphantom {{t_{o-o}^2} 
\tilde\alpha }} \right. \kern-\nulldelimiterspace} \tilde\alpha }$),
\cite{maezono98-2}
where $t_{o-o}$ ($t_{o-u}$) are the transfer integral between the
nearest-neighboring occupied/occupied (occupied/unoccupied) orbitals.
Orbital $F$ (0,0) leads to $t_{o-o}^{x,y}>t_{o-u}^{x,y}$
and thus the intra-plane bonds favor spin-$AF$ for our choice of the
parameters $\tilde\alpha \approx \tilde\beta$.
This destabilizes the spin $A$ structure in (100)-direction.
As the doping $x$ increases, the double-exchange interaction,
becomes more and more important.
This stabilizes the ferromagnetic bond within the plane, and 
$C_{x\left(y\right)}$ becomes positive.
\par
\begin{figure}[htbp]
\begin{center}
\vspace{0mm}
\hspace{0mm}
\epsfxsize=8cm
\epsfbox{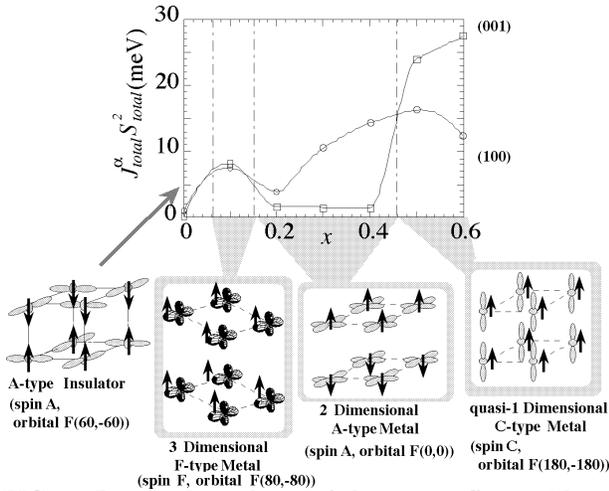}
\vspace{0mm}
\caption[aaa]{Doping-dependence of the spin stiffness.
The orbital and the spin structure are optimized at each point.
The enhancement of the spin-stiffness and the cross-over of 
the dimensionality are seen with increasing $x$. \cite{maezono99-1}}
\label{fig : F4}
\end{center}
\end{figure}
Fig. \ref{fig : F4} shows the {\it static} spin wave stiffness
as a function of the doping concentration $x$, including the contribution
from $t_{2g}$ ($J_S$).
We could reproduce the qualitative feature of the dimensional crossover
and the enhancement \cite{endoh97} of the stiffness in terms of the crossover
from the super- (for smaller $x$) to the double-exchange interactions 
(for larger $x$) accompanied with the change in the orbital ordering.
As $x$ increases, the spin structure changes from
spin $A$ insulator at $x=0$ into the nearly isotropic spin $F$ metal, 
to the spin $A$ metal with two-dimensional $d_{x^2-y^2}$-
orbital alignment, and to the spin $C$ metal with $d_{3z^2-r^2}$-
orbital. \cite{maezono98-1,maezono98-2}
Accordingly, the in-plane stiffness shows an increase,
moderately at the beginning and then rapidly in the region of the
spin $A$-metal. 
This reflects the fact that the double-exchange interaction is the most 
effective and prefers the $d_{x^2-y^2}$-orbital, i.e., the double-exchange 
interaction is basically {\it two-dimensional} with the $e_g$-orbitals.
In the spin $C$ metal for $x>0.4$, one-dimensional orbital along
(001)-direction gives rise to a steep increase of the stiffness
in this direction.
\par
The observed anisotropy of the spin stiffness is 
determined by the long range ordering of the orbitals.
Fig. \ref{fig : F4} also represents the cross-over of the 
dimensionality which we proposed in the previous report.
\cite{maezono98-2}
The stiffness changes from the nearly isotropic one 
in the spin $F$ state to the considerably strong two-dimensional 
one for spin $A$ metal, and 
to a quasi-one-dimensional one for spin $C$, reflecting the orbital 
transition with increasing $x$.
Yoshizawa $et\ al.$ \cite{yoshizawa98} observed such two-dimensional 
anisotropy of the stiffness for Nd$_{0.45}$Sr$_{0.55}$MnO$_3$, being
consistent with our result.
Quasi-one-dimensional anisotropy is predicted
for Nd$_{1-x}$Sr$_x$MnO$_3$ ($x>0.6$) \cite{kuwahara99,kajimoto99}.
\par
The in-plane spin stiffness $J_{\rm total}^{x\left(y\right)} S^2_{\rm total}$
in Fig. \ref{fig : F4} could be compared with the experiments.
In La$_{1-x}$Sr$_x$MnO$_3$, Endoh $et\ al.$\cite{endoh97}
observed the plateau of the velocity $v_x$ in the orbital-ordered
insulating state up to $x\sim0.12$ and then the velocity increases
in the spin $F$ metallic phase.
Comparing this with the calculation above, it seems that the 
moderate increase up to $x\sim0.15$
in Fig. \ref{fig : F4} corresponds to the plateau, while
the rapid increase for $x>0.15$ to the increasing velocity
observed by Endoh.\cite{endoh97}
Then orbital-ordered spin $F$ metallic state in Fig. \ref{fig : F4}
corresponds to the insulating spin $F$ phase in experiments.
Both the spin $F$- and $A$-metal in experiments, on the other hand,
seems to corresponds to the spin $A$-metal with $d_{x^2-y^2}$ orbital ordering
in the calculation.
This fits well orbital liquid picture by Ishihara {\it et al.}
\cite{ishihara97};
In the perfectly cubic system the orbital state in spin $F$ metal 
is described as the resonance among $d_{x^2-y^2}$, $d_{y^2-z^2}$, 
and $d_{z^2-x^2}$.
In the actual CMR compound, however, the slight lattice distortion 
\cite{fernandez98,martin96} may breaks the cubic symmetry to stabilize 
$d_{x^2-y^2}$, though it is still accompanied with large 
fluctuation around it.
\par
Now we turn to the absolute value of the stiffness in the spin $F$ metallic 
phase.
Taking the reported lattice constant and the magnitude of spin moment as,
$S^*=3/2+1/2(1-x)$, the experimental values of the \it{static} \rm spin
stiffness, $J_{\rm total}^{x} S^2_{\rm total}$, are
11.61 meV for La$_{0.7}$Sr$_{0.3}$MnO$_3$ \cite{martin96}
and 10.24 meV for Nd$_{0.7}$Sr$_{0.3}$MnO$_3$ \cite{fernandez98}, 
respectively.
These are in quite well coincidence with $J_{\rm total}^x S^2_{\rm total}$
=10.53 meV, estimated by RPA here with $x=0.3$, $d_{x^2-y^2}$-orbital ordering.
A simple tight-binding estimation of the static spin stiffness,
\begin{equation}
D=\frac{S^*}{2} \frac{\partial}{\partial(q^2)}
\sum_{\langle ij \rangle, \sigma} {t_{ij} \langle 0| 
c^\dagger_{i \sigma}
c_{j \sigma}}|0\rangle \ ,
\end{equation}
with $d_{x^2-y^2}$-orbital also gives the similar value, $\sim$
10 meV (with $t_0=0.72$ eV, $x=0.3$),
where the strong Coulombic interactions are reflected as the full orbital 
polarization $d_{x^2-y^2}$
(superexchange interactions are ignored).
This agreement can be understood in terms of the above orbital liquid 
picture as follows:
While the large orbital fluctuation around $d_{x^2-y^2}$ may cause the 
several anomalous behaviors in the transport properties,
it is not reflected to the stiffness constant because
the correction due to such a fluctuation has the wave vector dependence as
$\sim \left(1-\cos{q_\alpha}\right)^2$ \cite{nagaosa99}
as described in the next paragraph, doing little around
$\vec q =0$ and hence the stiffness constant.
Therefore the $d_{x^2-y^2}$-orbital ordering can give a good estimation
of the stiffness constant of spin $F$ metallic phase with a large orbital 
fluctuation.
\par
The softening observed near the zone boundary of the spin wave excitation 
can be understood in terms of the orbital fluctuation. \cite{nagaosa99}
When the normal vector of the resonating planer orbitals, $d_{x^2-y^2}$, 
$d_{y^2-z^2}$, and $d_{z^2-x^2}$, points along some bond direction, the
ferromagnetic double-exchange interaction disappears along this bond 
resulting, instead, the $AF$ interaction due to $t_{2g}$ orbitals.
Such an interaction between the orbital fluctuation and spin degrees of
freedom leads to, in the lowest order, the self-energy correction with 
the $k$-dependence as $\sim \left(1-\cos{q_\alpha}\right)^2$ for $(0,0,\xi)$- 
and $(0,\xi,\xi)$-direction but no (canceled out) correction for 
$(\xi,\xi,\xi)$ \cite{khaliullin99}, being consistent with the 
experiments \cite{hwang98}.
\par
%
%
The important implication concluded from the agreement
between the experimental and RPA-estimated value of the stiffness constant
is little influence of the JT polaron, 
\cite{millis95,roder96,millis96,alex99} at least on the spin dynamics.
JT polaron should reduce the double-exchange interaction in the doped region 
via a bandwidth reduction.
To describe this polaronic effect, we introduce here a generic
model;
Assume that the orbital configuration is relaxed to its stable 
one when the electron is occupying the site $i$.
We express the polaronic degrees of freedom by the bosons.
Now the electron operators $d^\dagger,d$ have no orbital index,
because of the sufficient orbital polarization,
\begin{eqnarray}
H &=& \sum_{ij,\sigma} t_{ij} d^\dagger_{i \sigma} d_{j \sigma}
     +  \sum_{i,\sigma} \sum_q g_q ( b_q + b^\dagger_{-q}) 
d^\dagger_{i \sigma} d_{i \sigma}
\nonumber \\
     &+& \sum_{k} \omega_k b_k^\dagger b_k
    + U \sum_i n_{i \uparrow} n_{i \downarrow}\ .
\end{eqnarray} 
This is the usual polaron Hamiltonian, and the following unitary 
transformation $\tilde U$ eliminates the coupling terms  between electrons 
and bosons,
\begin{equation}
\tilde U = \exp \biggl[ \sum_{i,\sigma}\sum_q 
\biggl( { { g_q} \over {\omega_q} } \biggr)
n_{i \sigma} e^{ i q \cdot R_i} ( b_q - b_{-q}^\dagger ) \biggr]
\ .
\end{equation}         
In terms of this $\tilde U$, the Hamiltonian $H$ is transformed as
\cite{mahan90}
\begin{eqnarray}
{\tilde H} &=& \tilde U^\dagger H \tilde U 
\nonumber \\
&=&\sum_{ij,\sigma}t_{ij} X^\dagger_i X_j d_{i \sigma}^\dagger d_{j \sigma}
 + \sum_q \omega_q b^\dagger_q b_q  \cr
& & \ \ \ \ \ - \sum_{i \sigma} \Delta \cdot n_{i \sigma}
 + U \sum_i n_{i \uparrow} n_{i \downarrow} \ ,
\end{eqnarray}
where $X_i = \exp[ \sum_q e^{ i q \cdot R_i} 
( g_q/\omega_q)(b_q - b^\dagger_{-q})]$, and
$\Delta = \sum_q g_q^2/\omega_q$ is the relaxation energy.  
We now derive the exchange interaction between spins in terms of the
perturbative expansion in $t_{ij}$.
The double-exchange interaction is the first order in $t_{ij}$, and
is reduced by the factor of 
$< X^\dagger_i X_j > = \exp[ - \sum_q | u_q |^2/2]$
($ u_q = (g_q/\omega_q)
( e^{i q \cdot R_i } -  e^{i q \cdot R_j } )$), which is exponentially small
when $g_q/\omega_q$ is large. 
This factor is nothing but the bandwidth reduction factor due to 
the polaronic effect.
On the other hand, for $x=0$, the superexchange interaction under the 
coupling with the polaron is given by,
\begin{equation}
J \!=\! 4 |t_{ij}|^2 \! \int_0^\beta \! d \tau G_0^2(\tau)\! 
\left<
\! X^\dagger_i(\tau) X_j(\tau) X^\dagger_j(0) X_i(0) \!\right>
\ ,
\end{equation}
where $G_0(\tau) = e^{- U \tau/2}$ is the Green's function 
for localized electrons.
Because we are interested in the large $U$ case, 
the integral is determined by the small $\tau$ region where
$\left< \! X^\dagger_i(\tau) X_j(\tau) X^\dagger_j(0) X_i(0) \!\right>
\cong e^{ - { \tilde \Delta} \tau }$ 
($ { \tilde \Delta } = \sum_q \omega_q |u_q|^2 $).
Then the polaronic effect is to replace $U$ by $U + {\tilde \Delta}$
in the expression for $J$, which is a minor correction when 
$U>> {\tilde \Delta}$ \cite{kugel81}, being in sharp contrast to the 
double-exchange interaction discussed above.
Polaronic effect should therefore correct the RPA-estimation of the
stiffness-enhancement as $x$ increases to be smaller.
Agreement between the observed and estimated stiffness in the doped 
region implies therefore that the spin dynamics is not so affected by
the JT polaron.
This is also pointed out by Quijada \it{et al}\rm.\cite{quijada98}
\par
Because the estimation is made under the assumption that the orbital
is almost fully polarized to $d_{x^2-y^2}$, the agreement also suggests
the large orbital polarization.
With the absence of the orbital polarization, the stiffness enhancement
should scale to electron density $(1-x)$ rather than the hole $x$.
The observed stiffness enhancement with increasing $x$ even in the metallic
region therefore also supports the large orbital polarization.
\par
\section{Conclusions}
We discussed the zero-temperature phase diagram and the spin dynamics
of the CMR compounds based on the model with a large orbital polarization.
The topology of the magnetic transition depending on the doping concentration
cannot be reproduced without a large orbital polarization.
This is because the double-exchange interaction is the most 
effective and prefers the $d_{x^2-y^2}$-orbital, i.e., the double-exchange 
interaction is basically {\it two-dimensional} with the orbital
polarization.
As for the ferromagnetic metallic phase the large orbital polarization
recovers the isotropy of the transport by forming a liquid state, i.e.,
the resonance among $d_{x^2-y^2}$, $d_{y^2-z^2}$ and $d_{z^2-x^2}$.
Spin $A$ phase seen in the moderately doped region has a stability against 
the canting with infinitesimal angle deviation from $\pi$, being in sharp 
contrast to the spin $A$ insulator.
Though it cannot be infinitesimal, finite canting angle between 
0 (spin $F$) and $\pi$ (spin $A$) can realize only if the orbital
is planer both in spin $F$ and $A$.
The observed metallic canting in 113-compounds is therefore an evidence that 
the ferromagnetic metallic phase consists of such a planer orbital.
The dispersion of the spin wave excitation evaluated in the RPA
is well fitted by the cosine curve.
This implies that the excitation is almost dominated by the 
nearest-neighbor exchange interaction even in the double-exchange
regime, being in conflict with the first principle result.
Estimated stiffness constant shows good agreement with the observed
values for the metallic region.
This strongly implies the absence of the JT-polaronic influence on
the spin dynamics in the doped region.
Based on the above orbital liquid picture, we could explain
the spin wave softening near the zone boundary, its anisotropy,
and no influence due to the orbital fluctuation on the stiffness 
constant.
\section{Acknowledgement}
The authors would like to thank S. Ishihara, S. Maekawa,
K. Hirota, Y. Endoh, I. Solovyev, K. Terakura, T. Kimura,
H. Kuwahara, R. Kajimoto, H. Yoshizawa, T. Akimoto, Y. Moritomo,
D. Khomskii, A.J. Millis, E.W. Plummer, J.F. Mitchell and Y. Tokura 
for their valuable discussions.
This work was supported by Priority Areas Grants from the Ministry
of Education, Science, Sports and Culture of Japan.
R.M. is supported by Research Fellowships of the 
Japan Society for the Promotion of Science (JSPS) for
Young Scientists.
%
%

%

\begin{references}
\bibitem{mitchell99}
Recent experiments with $x>0.5$ confirms that this prediction of the 
global topology is consistent with the observation, except some 
instability of the incommensulate phase seen for $x\sim 0.8$
(J.F. Mitchell, private communication, 1999).
\end{references}
\end{document}